\documentclass[a4paper,12pt]{article}
\usepackage{graphicx}
\usepackage{bm}
\usepackage{authblk}
\usepackage{xcolor}
\usepackage{subcaption}
\usepackage{amsmath}
\usepackage{amsfonts}
\usepackage[normalem]{ulem}
\usepackage{setspace}
\usepackage[margin=1in]{geometry}
\usepackage{tikz}
\usepackage{cite}
\usepackage{url}
\usepackage[font=small]{caption}

\title{Predicting ultrafast nonlinear dynamics in fibre optics with a recurrent neural network}
\date{}
 
\author[1,*]{\small Lauri Salmela}
\author[2]{\small Nikolaos Tsipinakis}
\author[2]{\small Alessandro Foi}
\author[3]{\small Cyril Billet}
\author[3]{\small John M. Dudley}
\author[1,*]{\small Go{\"e}ry Genty}

\affil[1]{\footnotesize Photonics Laboratory, Physics Unit, Tampere University, 33014 Tampere, Finland}
\affil[2]{\footnotesize Laboratory of Signal Processing, Tampere University, 33014 Tampere, Finland}
\affil[3]{\footnotesize Institut FEMTO-ST, Universit\'{e} Bourgogne Franche-Comt\'{e} CNRS UMR 6174, 25000 Besan\c{c}on, France}
\affil[*]{goery.genty@tuni.fi}

\begin{document}
 
\maketitle


\begin{abstract}
The propagation of ultrashort pulses in optical fibre displays complex nonlinear dynamics that find important applications in fields such as high power pulse compression and broadband supercontinuum generation. Such nonlinear evolution however, depends sensitively on both the input pulse and fibre characteristics, and optimizing propagation for application purposes requires extensive numerical simulations based on generalizations of a nonlinear Schr\"odinger-type equation.  This is computationally-demanding and creates a severe bottleneck in using numerical techniques to design and optimize experiments in real-time. Here, we present a solution to this problem using a machine-learning based paradigm to predict complex nonlinear propagation in optical fibres with a recurrent neural network, bypassing the need for direct numerical solution of a governing propagation model. Specifically, we show how a recurrent neural network with long short-term memory  accurately predicts the temporal and spectral evolution of higher-order soliton compression and supercontinuum generation, solely from a given transform-limited input pulse intensity profile. Comparison with experiments for the  case of soliton compression shows remarkable agreement in both  temporal and spectral domains. In optics, our results apply readily to the optimization of pulse compression and broadband light sources, and more generally in physics, they open up new perspectives for studies in all nonlinear Schr\"odinger-type systems in studies of Bose-Einstein condensates, plasma physics, and hydrodynamics. 
\end{abstract}

\newpage

 
\section{Introduction}

The past decade has seen major developments in the field of machine learning, and societal applications in  health-care, autonomous vehicles, and language processing are becoming commonplace \cite{jordan2015machine}. The impact of machine learning on basic research has been just as significant, and the use of advanced algorithmic tools in data analysis has resulted in new insights into many areas of science. In physics, there has been particular interest applying the tools of machine learning to study dynamical complex systems which evolve in time. These systems exhibit extreme sensitivity to small variations of the governing parameters, and the use of conventional numerical methods to understand and potentially control these dynamics is challenging.  

Nonlinear pulse propagation in optical fibre waveguides is known to exhibit highly complex evolution, and machine learning methods have been applied in a variety of ways to both optimize and analyze their spectrum or temporal intensity profile at the fibre output. For example from a feedback and control perspective, evolutionary algorithms (which are typically slow to converge) have been used in experiments optimizing particular characteristics of supercontinuum sources \cite{wetzel2018customizing, michaeli2018genetic}, as well as the experimental control of mode-locked fibre lasers \cite{andral2015fiber,pu2019intelligent,dudleymeng2020,kokhanovskiy2019machine2}.  Machine learning using neural networks has also been applied to classify experimentally different regimes of nonlinear propagation in modulation instability experiments \cite{narhi2018machine} or to determine the duration of short pulses from a fibre laser \cite{kokhanovskiy2019machine}. Applications to the control of mode-locking \cite{baumeister2018,kokhanovskiy2019machine} and pulse shaping \cite{finot2018nonlinear} have also been demonstrated numerically. Yet, all these applications have been restricted either to (slow) genetic algorithms or to feed-forward neural network architectures limited to determine the correspondence between a given input and some single output parameter. 

More generally, experiments in optical fibres are of very wide interest in nonlinear science since they provide a convenient means of studying nonlinear dynamics common to many nonlinear Schr\"{o}dinger equation (NLSE) systems including hydrodynamics, plasmas, and Bose-Einstein condensates.  However, because propagation in an NLSE system depends sensitively on both the input pulse and fibre characteristics, the design and analysis of experiments require extensive numerical simulations based on the numerical integration of the NLSE or its extensions. This is computationally-demanding and creates a severe bottleneck in using numerical techniques to design or optimize experiments in real-time. 

In this paper, we present a solution to this problem using machine-learning to predict complex nonlinear propagation in optical fibres with a recurrent neural network, bypassing the need for direct numerical solution of a governing propagation model.  The general context of our work is the recent development of machine learning approaches exploiting  \emph{knowledge-based} and \emph{model-free} methods to forecast and thus control complex evolving dynamics. Knowledge-based (or physics-informed) methods rely on some a priori knowledge of the mathematical model governing the physical system, and they perform especially well in capturing nonlinear dynamics \cite{brunton2016discovering,raissi2018deep, raissi2019physics}. In contrast, model-free forecasting is a purely \emph{data-driven} approach where a neural-network structure will learn the system dynamical behavior from a set of training data, without any prior knowledge of the physics of the system or any underlying governing equation(s). Model-free methods have been particularly successful in forecasting spatio-temporal dynamics of physical systems exhibiting high-dimensional chaos, instabilities and turbulence \cite{vlachas2018data, vlachas2019forecasting, pandey2020reservoir}, as well as reproducing the propagation dynamics of certain analytical solutions of the NLSE \cite{jiang2019model}.  

Our objective here is to significantly extend the use of model-free methods in nonlinear physics by showing how a long short-term memory (LSTM) recurrent neural network can fully reproduce the complex nonlinear dynamics of ultrashort pulse evolution in optical fibre governed by an NLSE system.  We study two particular cases of practical importance: high power pulse compression associated with the generation of Peregrine-soliton structures, and broadband optical supercontinuum generation.  In the first case, we  show how the network accurately predicts the temporal and spectral evolution of higher-order solitons and the appearance of the Peregrine soliton from a transform-limited intensity profile, and we also show how the predicted results agree with reported experimental measurements \cite{tikan2017universality}. We then expand our analysis to even more complex dynamics and show how the network can also predict the full development of an octave-spanning supercontinuum with fine details in the spectral and temporal domains. These results represent a significant extension of model-free methods applied to nonlinear optics, with potential important impact for high-field physics, nonlinear spectroscopy, and precision frequency comb metrology.  Moreover, we anticipate that our results will stimulate similar studies in all areas of physics where NLSE-like dynamics play a governing role.  



\section{Model-free modeling of nonlinear propagation dynamics}

The propagation of light in an optical fibre can be represented as a sequence of electric field complex amplitude distributions (spectral or temporal) at different points along the propagation path in the fibre. The amplitude at any specific propagation distance is naturally determined by the evolution which precedes it, and modelling this evolution is conventionally carried out by numerically integrating a governing NLSE model over a large number of elementary steps \cite{agrawal}. Unfortunately, this conventional approach can be extremely time-consuming.

Here, we show that such a direct numerical approach can in fact be replaced with model-free forecasting using a recurrent neural network (RNN). RNNs are a particular class of neural network that possess internal memory, allowing them to account for long-term dependencies and thus to robustly identify patterns in sequential data \cite{Lipton2015RNN}. The fact that RNNs intrinsically allow modelling of dynamic behavior makes them particularly adapted to the processing and predictions of time-series with applications in speech recognition, predictive texting, handwriting recognition, natural language processing, or stock market analysis. And they are equally a natural choice to predict the evolution of nonlinear propagation dynamics as a high power optical field propagates in an optical fibre.   

The particular form of RNN we use is the long short-term memory (LSTM) cell architecture \cite{hochreiter1997long}. Although other approaches such as reservoir computing or the gated recurrent unit would also be possible, our choice of LSTM is based on its simplicity of implementation and demonstrated success in various applications \cite{Carleoreview2019,reviewRNN2019}.  We train the network to be able to separately and independently forecast the evolution of temporal and spectral intensity during nonlinear pulse propagation in optical fibre, based only on the initial condition of a transform-limited pulse. Of course physically, the temporal and spectral field characteristics are tightly coupled, and it is therefore remarkable that the network is able to learn independently the temporal and spectral evolution dynamics using only intensity data.  In order to teach the network the pulse propagation dynamics, initial training is performed using ensembles of temporal and spectral intensity evolution maps, generated numerically using simulations of the NLSE (or its generalized version the GNLSE) for a range of input pulse characteristics.  In order to reduce the computational load during training, the simulation profiles are downsampled along both the propagation direction, and the temporal and spectral dimensions (see Methods).  


\begin{figure}[!ht]
  \centering
  \includegraphics[width=\linewidth]{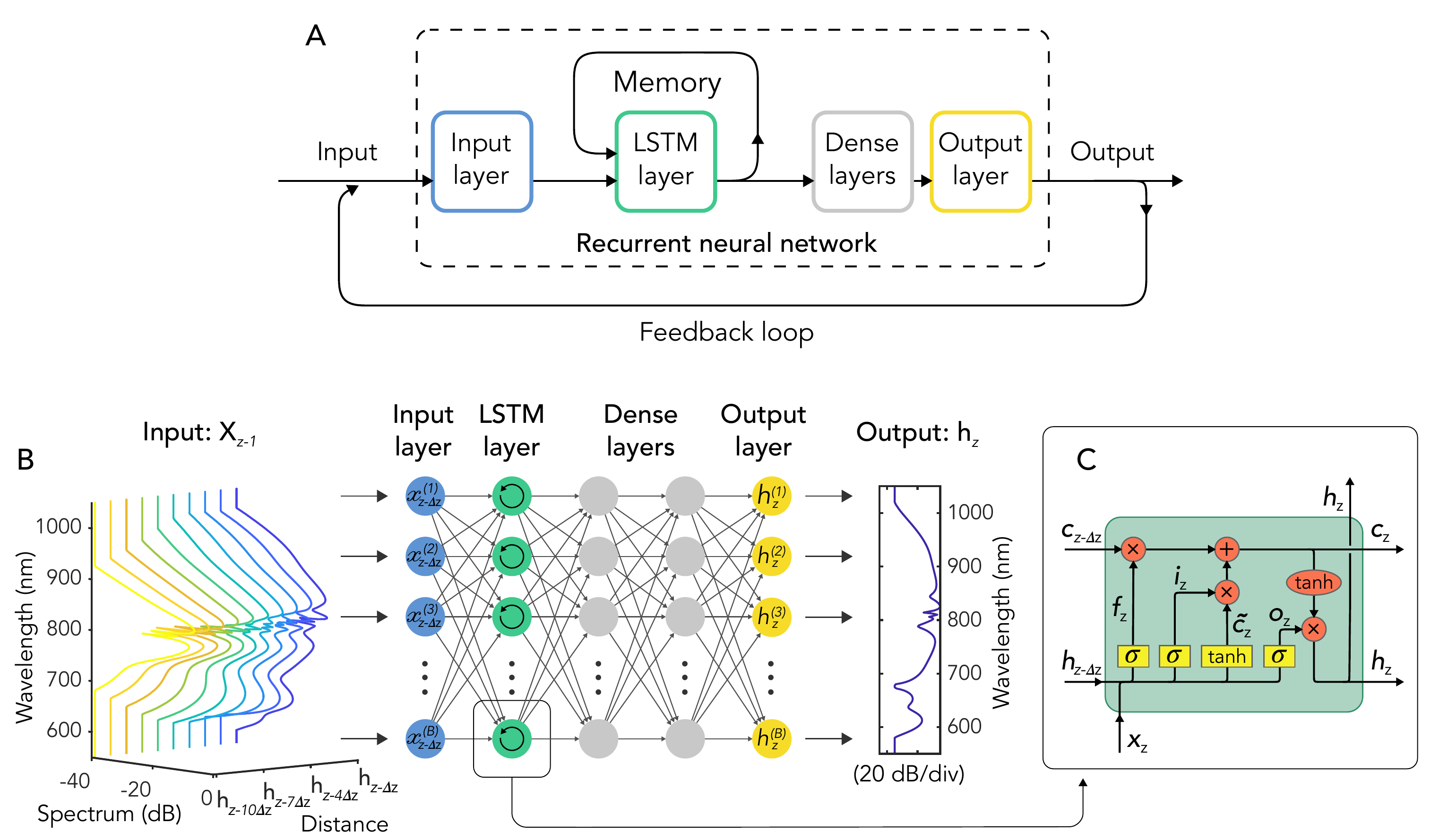}
\caption{(a) Schematic of the recurrent neural network architecture used showing: the input layer, the long short-term memory (LSTM) recurrent layer, two hidden (dense) layers, and the output layer. (b) The neural network uses the spectral (or temporal) intensity profiles $\mathbf{X}_{z-1}$ from the ten previous intensity profiles $h_{z-10\Delta z}..h_{z-\Delta z}$ in the evolution to yield the subsequent spectrum $h_{z}$. Each intensity profiles $h$ consists of $B$ intensity bins denoted as $x^{k}$ where $k$ indicates the bin number. (c) The LSTM cell receives the cell input, hidden and cell states from the previous step as an input, and output of the cell is the new hidden state that is also passed on to the next prediction step along with the new cell state. $x_i$ is the cell input where $i = z, z - \Delta z$, $h_i$ denotes the hidden state and $c_i$ the cell state. The yellow rectangles denote layer operations and the orange circles denote pointwise operations. See Methods for more details on the number of nodes used per layer, activation functions etc. More details and definition of the different cell elements are given in the Methods.} 
\label{fig:schmatic}
\end{figure}

A general schematic of the RNN is shown in Fig. \ref{fig:schmatic}(a) and an illustration of the training stage is shown in Fig. \ref{fig:schmatic}(b). Ten consecutive temporal or spectral intensity profiles $h_{z-10\Delta z}..h_{z-\Delta z}$ (i.e. the evolution from distance $z-10\Delta z$ to $z-\Delta z$) are fed to the RNN. Here $\Delta z$ corresponds to the sampling distance along the propagation direction (see Methods). The choice to feed the network with ten consecutive intensity profiles at propagation interval $\Delta z$ was found to be a good heuristic compromise between speed and performance (see Methods). These intensity profiles are then passed to the LSTM layer consisting of cells (Fig. \ref{fig:schmatic}(c)) governed by a specific algorithm (see Methods). Essentially, the LSTM layer uses 3 different types of information to predict the (spectral or temporal) intensity profile $h_{z}$ at distance $z$: (i) the intensity profile $h_{z-\Delta z}$ at distance $z-\Delta z$ which is the input of the LSTM layer, the hidden state of the layer corresponding to the predicted intensity profile $h_{z-2\Delta z}$ at distance $ z-2\Delta z$, and the cell state which contains the long-term dependency information from the intensity profiles $h_{z-10\Delta z}..h_{z-3\Delta z}$ corresponding to the evolution from distance $z-10\Delta z$ to $z-3\Delta z$.

The output of the LSTM layer is subsequently fed to a fully connected feed-forward neural network with two hidden (dense) layers whose function is to further improve the predicted intensity at distance $z$. The prediction made by the RNN (output layer) is compared with the intensity profile from the NLSE (or it generalized version GNLSE) simulations. The error is backpropagated to the weights and biases of the network nodes (both dense and LSTM layers) that are subsequently adjusted to minimize the prediction error. The RNN cycle is then initiated again with an updated input consisting of the consecutive temporal or spectral intensity profiles $h_{z-9\Delta z}..h_{z}$ till the full evolution is predicted. Note that the RNN loop is initiated with a ``cold start'' where the input sequence contains only the spectral or temporal intensity profile of pulses injected into the fibre (replicated ten times). In the prediction phase, the RNN model is tested using a separate set of temporal and spectral evolution data that was not used in the training phase.

\section{Results}
\label{sec:results}
\subsection{Higher-order soliton compression}
We begin by training the RNN to model the propagation of picosecond pulses in the anomalous dispersion regime of a highly nonlinear fibre. This propagation regime is of particular significance as it is associated with extreme self-focusing dynamics and practical ``higher-order soliton'' pulse compression schemes \cite{agrawal}. Moreover, the dynamics of this nonlinear temporal compression have been recently shown to be associated with the emergence of the celebrated Peregrine soliton that appears in the semiclassical limit of the NLSE \cite{tikan2017universality}.  

The training data was generated by performing 3,000 NLSE numerical simulations of propagation in 13~m of fibre using initial conditions of transform-limited hyperbolic-secant input pulses. The fibre parameters were kept constant between simulations and corresponded to experiments performed around 1550~nm \cite{tikan2017universality}.  On the other hand, we varied the pulse duration $\Delta\,\tau$ (FWHM) and peak power $P_0$  uniformly over the ranges 0.77--1.43~ps and 18.6--34.2 W, respectively.  This yields a variation in soliton number from $N=3.5-8.9$ where $N^2 = \gamma P_0 T_0^2/\beta_2$ with $\gamma$, $\beta_2$ the fibre nonlinear and group velocity dispersion parameters respectively, and $T_0 = \Delta\,\tau  /1.763 $.  See Methods for further details.   

\begin{figure}[!ht]
  \centering
  \includegraphics[width=\linewidth]{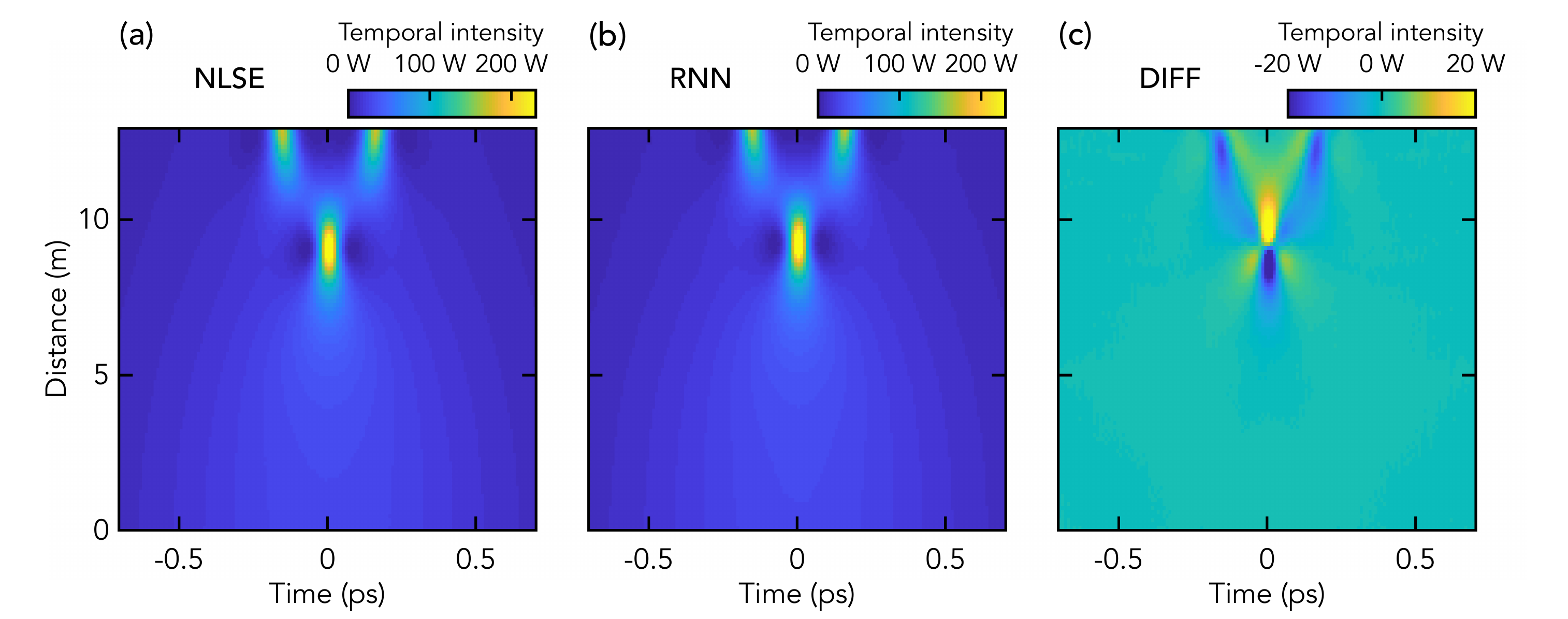}
\caption{Temporal intensity evolution of a 1.1~ps (full width at half maximum) pulse with 26.3~W peak power corresponding to an $N=6$ soliton injected into the anomalous dispersion regime of a 13~m long highly nonlinear fibre. The panels shows the result of NLSE numerical simulation (left), RNN prediction (middle), and relative difference (right). The RNN predictions use only the injected pulse intensity profile as input.}
\label{fig:pstime}
\end{figure}

We first illustrate the results obtained when training the network to model the temporal intensity evolution. Figure~\ref{fig:pstime} compares the evolution of the temporal intensity simulated using the NLSE (left panel) with that predicted by the RNN (central panel).  The particular results shown correspond to an input soliton number $N=6$.  One can see the overall excellent visual agreement between the propagation dynamic predicted by the RNN and those simulated from the NLSE. Also notice that the distance of maximum compression and associated temporal intensity profile is particularly well predicted by the RNN. The right panel shows the relative difference between the NLSE and RNN evolution maps, with a root mean square (RMS) error computed over the full evolution $R=0.04$ (see Methods). Comparisons between NLSE and RNN evolution for 100 different input condition spanning the full range of parameter variation showed similar results with a RMS error computed over the 100 evolution maps $R=0.097$ (see Methods).

\begin{figure}[!ht]
  \centering
  \includegraphics[width=\linewidth]{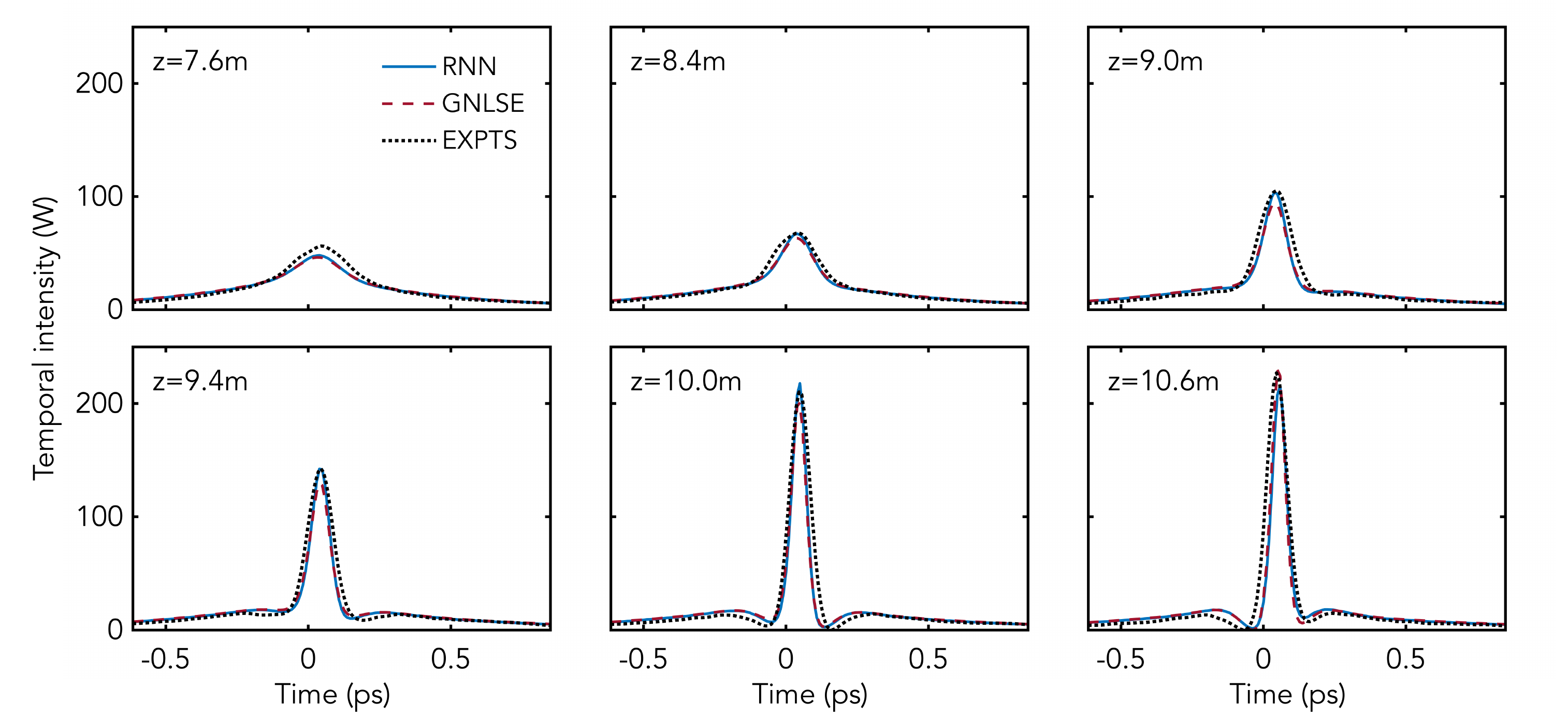}
\caption{Higher-order soliton ($N=6$) temporal intensity at selected distances predicted by the neural network (solid blue line), simulated with the NLSE (dashed red line), and experimentally measured (black dots). Experimental data from Ref. \cite{tikan2017universality}.}
\label{fig:psexpt}
\end{figure}

A more detailed comparison between the NLSE simulations, RNN prediction, and experimental measurements at selected distances is plotted in Fig.~\ref{fig:psexpt}. Note that in this case, third-order dispersion was also included in the training simulations (see Methods). The figure shows the intensity profiles predicted by the RNN (solid blue line), the profiles from the NLSE simulations (dashed red) as well as the experimental measurements (black dots) previously reported in \cite{tikan2017universality}. One can see remarkable agreement at all distances between the three sets of results, and we stress particularly that the RNN reproduces both the compressed central portion and the side lobes of the Peregrine soliton associated with maximal compression around 10~m.

\begin{figure}[!ht]
  \centering
  \includegraphics[width=\linewidth]{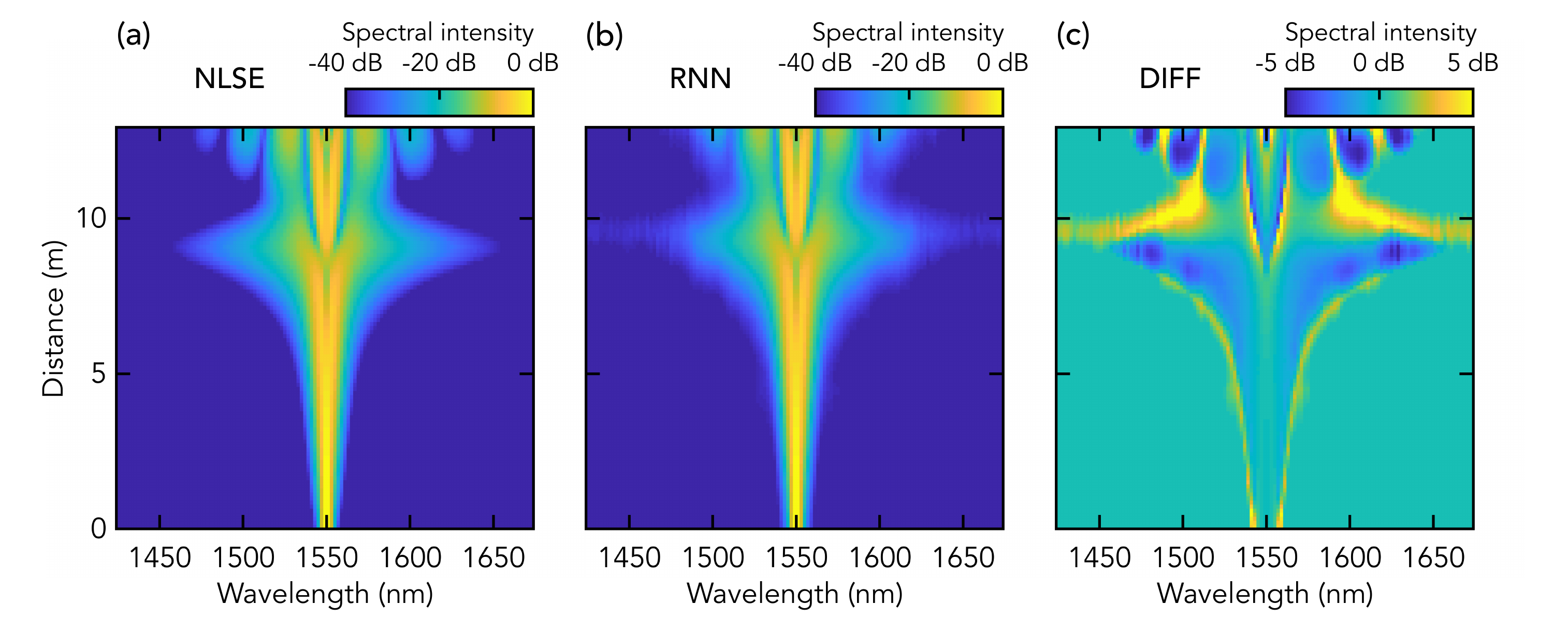}
\caption{Spectral intensity evolution of a 1.1~ps (full width at half maximum) pulse with 26.3~W peak power corresponding to an $N=6$ soliton injected into the anomalous dispersion regime of a 13~m long highly nonlinear fibre. The panels shows the result of numerical simulation (left), RNN prediction (middle), and relative difference (right). The RNN predictions use only the injected pulse spectrum as input.}
\label{fig:psspec}
\end{figure}

We also tested the ability of the RNN to predict the  propagation dynamics in the spectral domain from the corresponding input spectrum. Here we use the same ensemble of NLSE numerical simulations as for the temporal evolution, but this time we train the network by feeding the spectral intensity evolution. Results for input conditions identical to that of Figs.~\ref{fig:pstime} and \ref{fig:psexpt} are shown in Fig.~\ref{fig:psspec}. For convenient visualization, the evolution is plotted in logarithmic scale. The spectral evolution consists of an initial stage of spectral broadening dominated by self-phase modulation and corresponding to the compression observed in the time-domain. After the point of maximum expansion, we see a breathing phase of narrowing and re-expansion typical of higher-order soliton propagation. One can see excellent agreement between the dynamics predicted from the network and that simulated with the NLSE, with a relative discrepancy within a few dB over the entire evolution (RMS error computed over the full spectral evolution $R=0.106$). 

\begin{figure}[!ht]
  \centering
  \includegraphics[width=\linewidth]{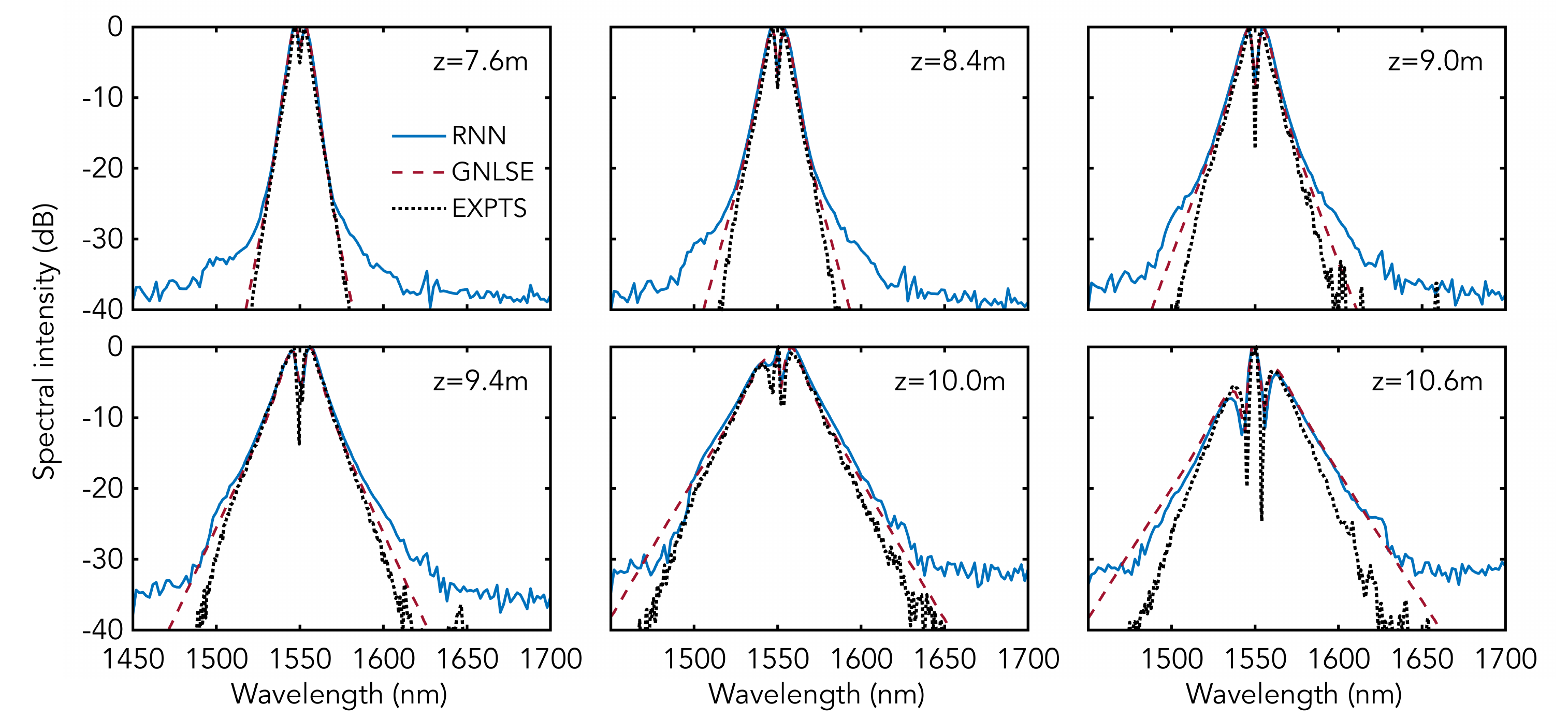}
\caption{Higher-order soliton ($N=6$) spectral intensity at selected distances predicted by the neural network (solid blue line), simulated with the NLSE (dashed red line), and experimentally measured (black dots).}
\label{fig:psexpt_spec}
\end{figure}

The excellent correspondence is confirmed in Fig.~\ref{fig:psexpt_spec} when plotting detailed comparison between the RNN predicted (blue), simulated (dashed red), and experimentally measured spectra (black dots) at selected distances around the maximal temporal compression point as previously considered and which is naturally also the point of maximum spectral broadening. In particular, one can see the excellent agreement between the NLSE and RNN results over a 25~dB dynamic range. We also performed a series of tests for 100 different input pulse spectra spanning the full range of parameter variation and found similar network performances in terms of predicted evolution with a RMS error $R=0.161$ (computed over the 100 evolution maps tested).



\subsection{Supercontinuum generation}
We next extended our study to even more complex propagation dynamics and the generation of a broadband supercontinuum. Here, we focus our attention to SC generated by injecting femtosecond pulses into the anomalous dispersion of a highly nonlinear fibre. This regime is of particular significance as it has been shown to be associated with high spectral coherence and the generation of stable frequency combs as well as to yield the broadest SC spectra \cite{dudley2006supercontinuum}. 

In order to test whether a recurrent neural network could learn SC generation dynamics and model their evolution, we generated an ensemble of SC propagation dynamics using the generalized NLSE (GNLSE) that includes the frequency-dependence of dispersion and nonlinearity, and the delayed Raman response \cite{dudley2006supercontinuum}. Specifically, we simulated the propagation of 100~fs transform-limited pulses at 810~nm injected into the anomalous dispersion regime of a 20~cm long photonic crystal fibre with zero-dispersion at 750~nm similar to that used in \cite{narhi2018machine}. See Methods for detailed parameter values. The ensemble includes simulations for a transform-limited input pulse with peak power uniformly distributed in the 500~W to 2~kW range that yields SC spectra with different characteristics, from isolated dispersive wave generation to fully developed octave-spanning SC with very fine spectral features. We emphasize that although the input pulse duration was kept constant for all the simulations, predicted results for other durations show similar agreement with the GNLSE as the specific cases discussed below.  

We begin by training the network from the temporal intensity evolution. Similarly to the higher-order soliton compression case, the simulation profiles are downsampled along both the propagation direction, and the temporal and spectral dimensions (see Methods). After training, the RNN model is tested for an input peak power not used in the training stage and the predicted evolution is compared with that directly simulated with the GNLSE for the same input power.  


\begin{figure}[!ht]
  \centering
  \includegraphics[width=\linewidth]{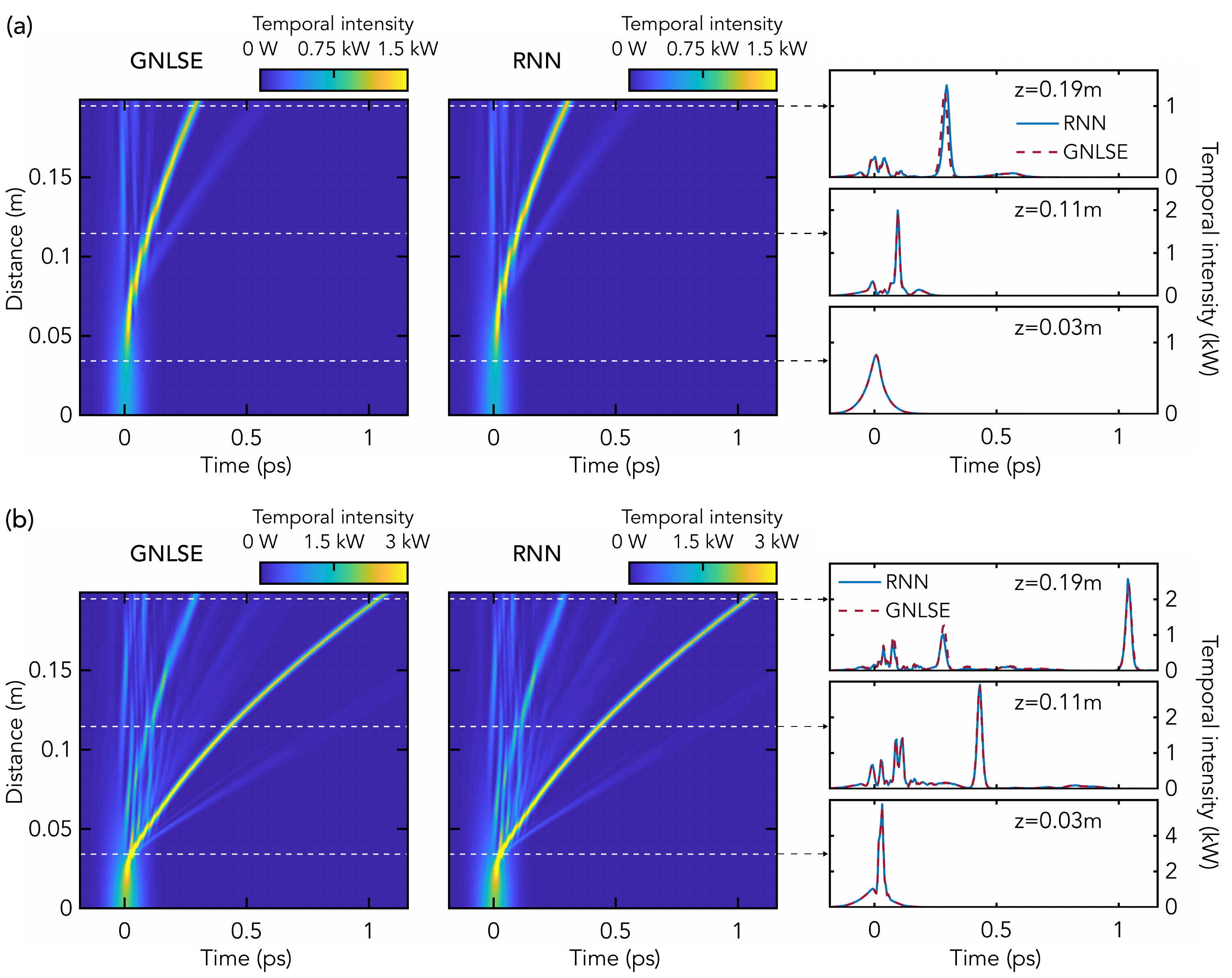}
\caption{Temporal evolution of supercontinuum. The left panel shows the
numerical simulation (GNLSE) of a supercontinuum evolution in a 20~cm photonic crystal fibre for 100~fs pulse with peak power of 630~W (a) and 1.96~kW (b). See Methods for a full description of the fibre parameters. The middle panel shows the predicted (RNN) temporal intensity evolution for the same initial temporal intensity profile as in the GNLSE simulations. The right panel shows the comparison between the predicted (solid blue line) and simulated (dashed red line) profiles at selected  distances indicated by white dashed lines.}
\label{fig:sc_time}
\end{figure}

Results are shown in Fig.~\ref{fig:sc_time}(a) and (b) for an input peak power of 630~W and 1.96~kW corresponding to an input soliton number of $N=4.6$ and $N=8.1$, respectively. These values were chosen as they lead to SC with very distinct characteristics. The left panel shows the temporal intensity evolution from the GNLSE simulation and the central panel shows the predicted evolution by the RNN. The SC generation process arises from soliton dynamics including higher-order soliton compression, soliton fission and dispersive waves emission on the short wavelength side \cite{dudley2006supercontinuum}. For longer propagation distances, solitons emerging from the fission experience the Raman self-frequency shift expanding the SC spectrum towards the long wavelengths side \cite{dudley2006supercontinuum}. Significantly, in both scenarios, one can see the excellent visual agreement between the GNLSE simulations and RNN model. The point of soliton fission and dispersive emission as well as the red-shifting solitons parabolic trajectories are perfectly reproduced by the network. Quantitatively, the relative difference remains within a few dBs over the entire evolution down to the -30~dB bandwidth. The RMS error calculated over the full intensity evolution is $R=0.097$ and $R=0.049$ for Figs.~\ref{fig:sc_time}(a) and (b), respectively. The remarkable ability of the RNN to predict very complex nonlinear dynamics is further highlighted in the right panel where we plot detailed comparison between the predicted and simulated SC temporal intensity at selected distances along the propagation where we can see how the amplitude and delay of the dispersive waves and Raman-shifted solitons are also predicted with excellent accuracy at all stages of the propagation. Additional predictions ran for 50 different values of pulse peak power (not used in the training phase) showed also very good agreement with the GNLSE simulations (RMS error $R=0.176$ computed over 50 different evolution maps tested).


\begin{figure}[!ht]
  \centering
  \includegraphics[width=\linewidth]{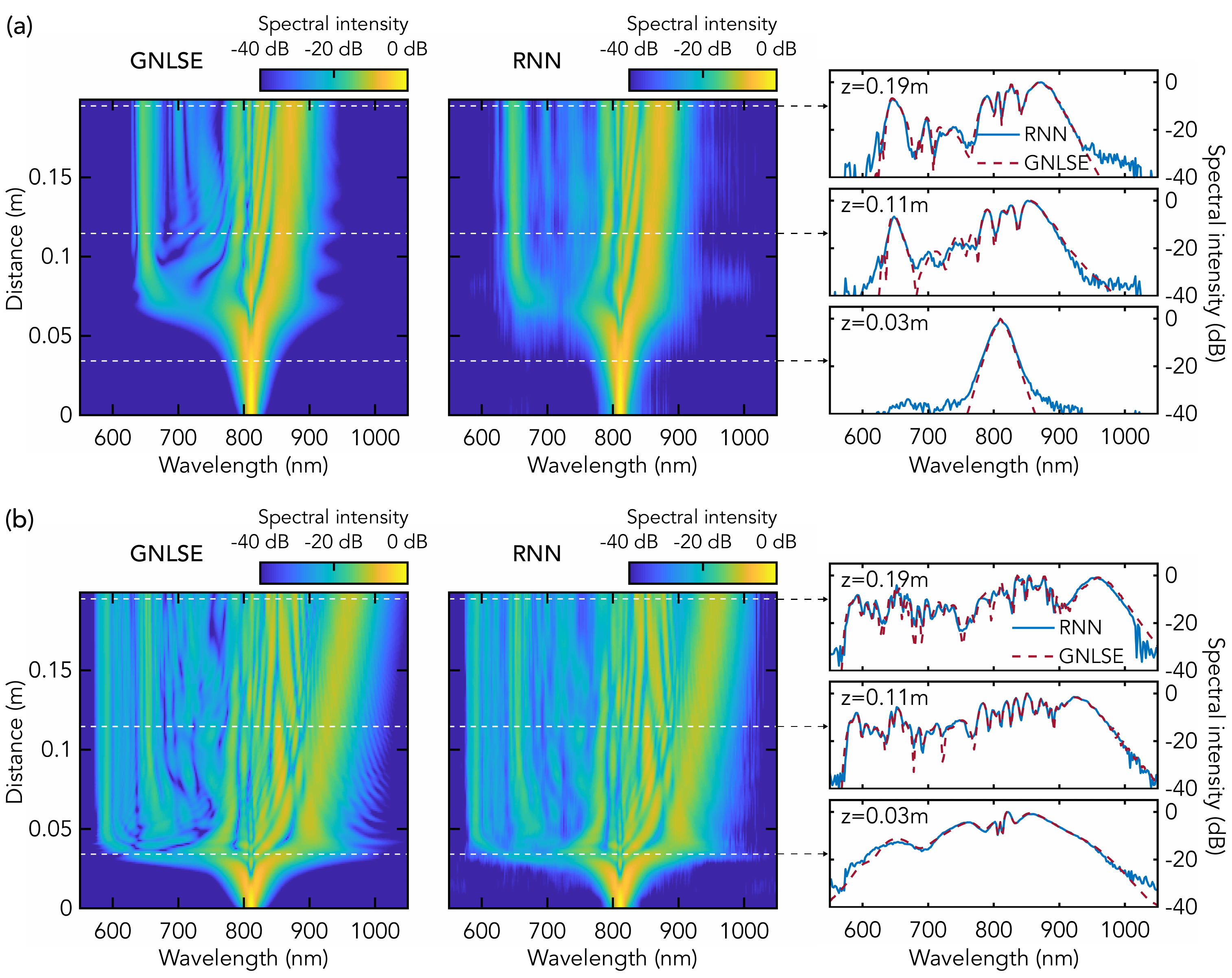}
\caption{Spectral evolution of supercontinuum. The left panel shows the
numerical simulation (GNLSE) of a supercontinuum evolution in a 20~cm photonic crystal fibre for 100~fs pulse with peak power of 630~W (a) and 1.96~kW (b). See Methods for a full descritpion of the fibre parameters. The middle panel shows the predicted (RNN) temporal intensity evolution for the same initial spectral intensity profile as in the GNLSE simulations. The right panel shows the comparison between the predicted (solid blue line) and simulated (dashed red line) profiles at selected  distances indicated by white dashed lines.}
\label{fig:sc}
\end{figure}

We then tested the ability of the RNN model to predict the SC spectral intensity evolution from the input pulse spectrum. The results for an input peak power of 630~W and 1.96~kW are shown in Fig.~\ref{fig:sc}(a) and (b), respectively. For convenient visualization, the evolution is plotted in logarithmic scale. In the case of lower peak power, one can see that the SC spectrum at the fibre output essentially consists of an isolated dispersive wave and solitons with a limited amount of red-shift. For larger input peak power, we see multiple dispersive wave emission and well-separated Raman-shifted solitons resulting in an octave-spanning SC.  Again, we can see very good visual agreement between the simulated and predicted evolution maps and that all spectral features including dispersive waves, Raman-shifted solitons and their interference that lead to fine spectral features are perfectly reproduced by the RNN. Additional predictions ran for 50 different input pulse peak power (not used in the training phase) showed also very good agreement with the GNLSE simulations (RMS error $R=0.12$).


\section{Conclusion}
We have shown that machine learning techniques can bring new insight into the study and prediction of nonlinear optical systems. Specifically, we have demonstrated that a recurrent neural network with long short-term memory can learn the complex dynamics associated with the nonlinear propagation of short pulses in optical fibres inclduing higher-order soliton compression and supercontinuum generation using solely the pulse intensity profile as input condition.  The network is also able to reproduce the dynamics both in the temporal and spectral domain, and for the particular case of higher-order soliton compression we have been able to confirm that the predicted evolutions maps are also in excellent agreement with experiments. Our results are particularly significant as applications of machine learning to ultrafast dynamics have previously been restricted to slow genetic algorithms or feed-forward neural networks designed to establish the transfer function between specific input-output parameters \cite{farfan2018femtosecond,finot2018nonlinear,andral2015fiber,kokhanovskiy2019machine,kokhanovskiy2019machine2}. 

From an application point of view, we expect that neural networks will very soon become an important and standard tool for analysing complex ultrafast dynamics, for optimizing the generation of broadband spectra and frequency combs, as well as for designing ultrafast optics experiments. Future steps may expand the parameter space of the RNN operation by including additional training variables as e.g. the nonlinear fibre parameters. The evolution prediction may be extended to the complex field (amplitude and phase), and one could also envisage to use reverse-engineering in order to optimize the pump pulse characteristics for the generation of on-demand temporal and spectral intensity profiles at the fibre (or waveguide) output. From a more fundamental perspective, we believe that the use of recurrent neural networks will impact on future design and analysis of nonlinear physics experiments as they represent a natural candidate for exploring and analyzing complex operation regimes with long-term dependencies.  

\section*{Acknowledgements}
LS acknowledges the Faculty of Engineering and Natural Sciences graduate school of Tampere University.
JD acknowledges the French Agence Nationale de la Recherche (ANR-15-IDEX- 0003, ANR-17-EURE-0002).
GG acknowledges the Academy of Finland (298463, 318082, Flagship PREIN 320165).
The authors also thank D. Brunner for useful discussions.

\section*{Methods}

\subsection*{Numerical simulations}
The numerical simulations in this work are from the NLSE and its generalized extension (1+1D) that describe the propagation of the slowly-varying optical field envelope.

\textbf{Higher-order soliton compression.}
We model the propagation of short pulses in the anomalous dispersion regime of a 13~m nonlinear optical fibre. The pulses have a hyperbolic-secant intensity profile centered at 1550~nm with pulse duration and peak power varying from 0.77 to 1.43~ps and from 18.41 to 34.19~W, respectively. The nonlinear coefficient of the fibre is $\gamma = 18.4 \times 10^{-3}$~W$^{-1}$m$^{-1}$, and the group-velocity dispersion coefficient at 1550~nm is $\beta_2 = -5.23 \times 10^{-27}$~s$^2$m$^{-1}$. When comparing with the experiments, third order dispersion ($\beta_3 = 4.27 \times 10^{-41}$~s$^3$m$^{-1}$) was also included in the training in addition to a small input pulse asymmetry caused by the experimental implementation \cite{tikan2017universality}. The simulations use 1024 spectral/temporal grid points with temporal window size of 10~ps and a step size of 0.13 mm (10,000 steps). For completeness, shot noise is added via one-photon-per-mode with random phase in the frequency domain, although noise effects were found to play no significant physical role in the regime of coherent propagation studied here.

\textbf{Supercontinuum generation.}
We model the propagation of sech-type pulse centered at 810~nm and with pulse duration of 100~fs. The peak power of the input pulse is randomly varied in range of 0.5-2~kW. The pulses are injected in the anomalous dispersion regime of a 20~cm nonlinear optical fibre, including higher-order dispersion terms, self-steepening and Raman effect. The nonlinear coefficient of the fibre is $\gamma = 0.1$~W$^{-1}$m$^{-1}$, and the Taylor-series expansion coefficients of the dispersion at 810~nm are 
$\beta_2 = -9.59 \times 10^{-27}$~s$^2$m$^{-1}$,
$\beta_3 = 7.84 \times 10^{-41}$~s$^3$m$^{-1}$,
$\beta_4 = -6.84 \times 10^{-56}$~s$^4$m$^{-1}$,
$\beta_5 = -4.78 \times 10^{-70}$~s$^5$m$^{-1}$,
$\beta_6 = 2.71 \times 10^{-84}$~s$^6$m$^{-1}$ and
$\beta_7 = -5.00 \times 10^{-99}$~s$^7$m$^{-1}$.
The simulations use 2048 spectral/temporal grid points with temporal window size of 5~ps and a step size of 0.02 mm (10,000 steps). Shot noise is added via one-photon-per-mode with random phase in the frequency domain, but in the coherent propagation regime studied here noise effects were found to play no significant physical role.

\subsection*{Recurrent neural networks}
\textbf{LSTM network operation}
The operation of an LSTM cell can be described at time step $t$ with input $\mathbf{x}_t \in \mathbb{R}^{d_o}$ by a set of equations given by \cite{hochreiter1997long}
\begin{equation}
  \begin{split}
    \mathbf{f}_t &= \sigma (\mathbf{W}_f[\mathbf{h}_{t-1}, \mathbf{x}_t] + \mathbf{b}_f)\\
    \mathbf{\tilde{c}}_t &= \text{tanh} (\mathbf{W}_c[\mathbf{h}_{t-1}, \mathbf{x}_t] + \mathbf{b}_c)\\
    \mathbf{o}_t &= \sigma (\mathbf{W}_o[\mathbf{h}_{t-1}, \mathbf{x}_t] + \mathbf{b}_o)
  \end{split}
\qquad
  \begin{split}
    \mathbf{i}_t &= \sigma (\mathbf{W}_i[\mathbf{h}_{t-1}, \mathbf{x}_t] + \mathbf{b}_i)\\
    \mathbf{c}_t &= \mathbf{f}_t \odot \mathbf{c}_{t-1} + \mathbf{i}_t \odot \mathbf{\tilde{c}}_t\\
    \mathbf{h}_t &= \mathbf{o}_t \odot \text{tanh}(\mathbf{c}_t),
  \end{split}
\end{equation}
where $\mathbf{f}_t$, $\mathbf{i}_t$ and $\mathbf{o}_t \in \mathbb{R}^{d_h}$ are the forget, input and output gate vectors, respectively, with $d_h$ denoting the dimensionality of the hidden state (i.e. the number of hidden units).
Vectors $\mathbf{c}_t$ and $\mathbf{h}_t \in \mathbb{R}^{d_h}$ are the updated cell and hidden state, respectively, and $\mathbf{W}_f$, $\mathbf{W}_i$, $\mathbf{W}_c$ and $\mathbf{W}_o \in \mathbb{R}^{d_h \times (d_h+d_o)}$ represent the cell weights and $\mathbf{b}_f$, $\mathbf{b}_i$, $\mathbf{b}_c$ and $\mathbf{b}_o \in \mathbb{R}^{d_h}$ are the biases. The sign $\odot$ denotes point-wise multiplication. The weights and biases of the network are iteratively trained via backpropagation \cite{werbos1990backpropagation}.

\textbf{Feed-forward network operation}
The operation of the fully-connected layers is similar to that in Ref.~\cite{narhi2018machine}. The codes were written in Python using Keras \cite{chollet2015keras} with Tensorflow backend \cite{abadi2016tensorflow}.

\textbf{Comparison between RNN prediction and (G)NLSE simulations}
A quantitative comparison between the network predicted evolution map and that simulated with the (G)NLSE can be performed using the average (normalized) root mean squared (RMS) errors as a metric:
\begin{equation}
    \label{eq:error}
    R = \sqrt{\frac{\sum_{i,d} (x_{m,i,d} - \hat{x}_{m,i,d})^2}{\sum_{i,d} (x_{m,i,d})^2}},
\end{equation}
where $\mathbf{x}_m$ and $\hat{\mathbf{x}}_m$ denote the (G)NLSE and RNN predicted intensity profile for realization $m$. The variables $i$ and $d$ indicate summation over the intensity profiles and propagation steps, respectively. When evaluating the performance of the prediction over an ensemble of $M$ evolution maps, the RMS error is calculated over $M$ distinct realizations.

\textbf{Higher-order soliton compression.}
An ensemble of 3,000 numerical simulations was generated. 2,900 realizations are used for the training of the RNN and 100 unseen realizations are used for testing. The simulated intensity evolution maps are uniformly downsampled at a constant propagation step of $\Delta z = 0.13~$m, yielding 101 intensity profiles along propagation for each simulated evolution map. At every of the 101 steps, the intensity profile is convolved and downsampled with a 10~fs full width at half maximum super-Gaussian temporal filter corresponding to 145 equally spaced bins in the $\left[-0.7,+0.7\right]$~ps time interval. The spectral intensity profiles are convolved and downsampled with a 2~nm full width at half maximum super-Gaussian spectral filter resulting in 126 equally spaced intensity bins spanning from 1425 to 1675~nm. The temporal and spectral intensity profiles are normalized by the peak intensity over all realizations. We emphasize that because from an experimental viewpoint intensity profiles (spectral or temporal) are more straightforward to measure that the full field, we choose to only use transform-limited intensity profiles during the RNN training while the phase information is completely omitted. Both for the temporal and spectral evolution, the network is trained with intensity profiles in linear scale.  When comparing with the experiments, to account for the slight input pulse asymmetry the NLSE simulated intensity profiles of every map used in the training phase of the RNN were convolved and downsampled with a 10~fs full width at half maximum super-Gaussian temporal filter corresponding to 151 equally spaced bins in the $\left[-0.62,+0.85\right]$~ps time interval. The spectral intensity profiles were convolved and downsampled similarly to the case of ideal higher-order soliton propagation but spanning from 1450 to 1700~nm.

The LSTM and two hidden layers consist of 161 nodes each with ReLU activations $f(x) = \text{max}(0,x)$, and the output layer consists of 151 and 126 nodes for temporal and spectral predictions, respectively, with sigmoid activation $f(x) = 1/(1+\text{exp}(-x)$. The network is trained for 60 and 120 epochs with RMSprop optimizer \cite{tieleman2012lecture} and adaptive learning rate for the temporal and spectral intensity predictions, respectively. 

The input of the RNN consists of ten consecutive temporal or spectral intensity profiles $h_{z-10\Delta z}..h_{z-\Delta z}$ at distance along the fibre $z-10\Delta z$ to $z-\Delta z$. 

A smaller number of intensity profiles was also found to give satisfactory results but of course this is at the expense of the relative prediction error which increases from 0.097 to 0.174 (temporal intensity evolution of higher-order soliton) when reducing the number of consecutive intensity profiles from ten to five. As the number of consecutive intensity profiles used in the training is increased, the training time also increases and therefore the training process is always a compromise between the prediction accuracy and the time required to train the network.

\textbf{Supercontinuum generation.}
An ensemble of 1,300 numerical simulations was generated. 1,250 realizations were used for training the RNN and 50 realizations for testing. The simulated intensity evolution maps are uniformly downsampled at a constant propagation step of $\Delta z = 0.2~$mm, yielding 200 intensity profiles along propagation for each simulated evolution. In order to reduce the computational load, when training the RNN to predict temporal intensity maps, the profiles at each of the 200 steps are convolved and downsampled with a 10~fs full width at half maximum super-Gaussian temporal filter corresponding to 276 equally spaced bins spanning in the $\left[-0.18,+1.16\right]$~ps time interval. Note that the asymmetry in the modeled time interval is implemented to account for the soliton self-frequency shift effect. When training the RNN from spectral intensity profiles, each  spectrum is convolved and downsampled with a 2~nm full width at half maximum super-Gaussian spectral filter such that the wavelength grid consisted of 251 spectral intensity bins spanning from 550 to 1050~nm. The profiles are normalized by the peak intensity over all realizations.

For the temporal intensity evolution, the LSTM and two hidden layers consist of 300 nodes each with ReLU activations, and the output layer consists of 276 nodes with sigmoid activation. The network was trained for 120 epochs with RMSprop optimizer and adaptive learning rate. For the spectral intensity evolution, the LSTM and two hidden layers consist of 250 nodes each with ReLU activations, and the output layer consists of 251 nodes with sigmoid activation. The network is trained for 100 epochs.

\bibliographystyle{ieeetr}
\bibliography{bibl}

\begin{thebibliography}{10}

\bibitem{jordan2015machine}
M.~I. Jordan and T.~M. Mitchell, ``Machine learning: Trends, perspectives, and
  prospects,'' {\em Science}, vol.~349, no.~6245, pp.~255--260, 2015.

\bibitem{wetzel2018customizing}
B.~Wetzel, M.~Kues, P.~Roztocki, C.~Reimer, P.-L. Godin, M.~Rowley, B.~E.
  Little, S.~T. Chu, E.~A. Viktorov, D.~J. Moss, A.~Pasquazi, M.~Peccianti, and
  R.~Morandotti, ``Customizing supercontinuum generation via on-chip adaptive
  temporal pulse-splitting,'' {\em Nature Communications}, vol.~9, no.~1,
  pp.~1--10, 2018.

\bibitem{michaeli2018genetic}
L.~Michaeli and A.~Bahabad, ``Genetic algorithm driven spectral shaping of
  supercontinuum radiation in a photonic crystal fiber,'' {\em Journal of
  Optics}, vol.~20, no.~5, p.~055501, 2018.

\bibitem{andral2015fiber}
U.~Andral, R.~S. Fodil, F.~Amrani, F.~Billard, E.~Hertz, and P.~Grelu, ``Fiber
  laser mode locked through an evolutionary algorithm,'' {\em Optica}, vol.~2,
  no.~4, pp.~275--278, 2015.

\bibitem{pu2019intelligent}
G.~Pu, L.~Yi, L.~Zhang, and W.~Hu, ``Intelligent programmable mode-locked fiber
  laser with a human-like algorithm,'' {\em Optica}, vol.~6, no.~3,
  pp.~362--369, 2019.

\bibitem{dudleymeng2020}
J.~M. Dudley and F.~Meng, ``Toward a self-driving ultrafast fiber laser,'' {\em
  {Light: Science \& Applications}}, vol.~9, p.~26, 2020.

\bibitem{kokhanovskiy2019machine2}
A.~Kokhanovskiy, A.~Ivanenko, S.~Kobtsev, S.~Smirnov, and S.~Turitsyn,
  ``Machine learning methods for control of fibre lasers with double gain
  nonlinear loop mirror,'' {\em Scientific Reports}, vol.~9, no.~1, p.~2916,
  2019.

\bibitem{narhi2018machine}
M.~N{\"a}rhi, L.~Salmela, J.~Toivonen, C.~Billet, J.~M. Dudley, and G.~Genty,
  ``Machine learning analysis of extreme events in optical fibre modulation
  instability,'' {\em Nature Communications}, vol.~9, no.~1, pp.~1--11, 2018.

\bibitem{kokhanovskiy2019machine}
A.~Kokhanovskiy, A.~Bednyakova, E.~Kuprikov, A.~Ivanenko, M.~Dyatlov,
  D.~Lotkov, S.~Kobtsev, and S.~Turitsyn, ``Machine learning-based pulse
  characterization in figure-eight mode-locked lasers,'' {\em Optics Letters},
  vol.~44, no.~13, pp.~3410--3413, 2019.

\bibitem{baumeister2018}
T.~Baumeister, S.~L. Brunton, and J.~N. Kutz, ``Deep learning and model
  predictive control for self-tuning mode-locked lasers,'' {\em Journal of the
  Optical Society of America B}, vol.~35, pp.~617--626, 2018.

\bibitem{finot2018nonlinear}
C.~Finot, I.~Gukov, K.~Hammani, and S.~Boscolo, ``Nonlinear sculpturing of
  optical pulses with normally dispersive fiber-based devices,'' {\em Optical
  Fiber Technology}, vol.~45, pp.~306--312, 2018.

\bibitem{brunton2016discovering}
S.~L. Brunton, J.~L. Proctor, and J.~N. Kutz, ``Discovering governing equations
  from data by sparse identification of nonlinear dynamical systems,'' {\em
  Proceedings of the National Academy of Sciences}, vol.~113, no.~15,
  pp.~3932--3937, 2016.

\bibitem{raissi2018deep}
M.~Raissi, ``Deep hidden physics models: Deep learning of nonlinear partial
  differential equations,'' {\em The Journal of Machine Learning Research},
  vol.~19, no.~1, pp.~932--955, 2018.

\bibitem{raissi2019physics}
M.~Raissi, P.~Perdikaris, and G.~E. Karniadakis, ``Physics-informed neural
  networks: A deep learning framework for solving forward and inverse problems
  involving nonlinear partial differential equations,'' {\em Journal of
  Computational Physics}, vol.~378, pp.~686--707, 2019.

\bibitem{vlachas2018data}
P.~R. Vlachas, W.~Byeon, Z.~Y. Wan, T.~P. Sapsis, and P.~Koumoutsakos,
  ``Data-driven forecasting of high-dimensional chaotic systems with long
  short-term memory networks,'' {\em Proceedings of the Royal Society A:
  Mathematical, Physical and Engineering Sciences}, vol.~474, no.~2213,
  p.~20170844, 2018.

\bibitem{vlachas2019forecasting}
P.~R. Vlachas, J.~Pathak, B.~R. Hunt, T.~P. Sapsis, M.~Girvan, E.~Ott, and
  P.~Koumoutsakos, ``Forecasting of spatio-temporal chaotic dynamics with
  recurrent neural networks: A comparative study of reservoir computing and
  backpropagation algorithms,'' {\em arXiv preprint arXiv:1910.05266}, 2019.

\bibitem{pandey2020reservoir}
S.~Pandey and J.~Schumacher, ``Reservoir computing model of two-dimensional
  turbulent convection,'' {\em arXiv preprint arXiv:2001.10280}, 2020.

\bibitem{jiang2019model}
J.~Jiang and Y.-C. Lai, ``Model-free prediction of spatiotemporal dynamical
  systems with recurrent neural networks: Role of network spectral radius,''
  {\em Physical Review Research}, vol.~1, no.~3, p.~033056, 2019.

\bibitem{tikan2017universality}
A.~Tikan, C.~Billet, G.~El, A.~Tovbis, M.~Bertola, T.~Sylvestre, F.~Gustave,
  S.~Randoux, G.~Genty, P.~Suret, and J.~M. Dudley, ``Universality of the
  peregrine soliton in the focusing dynamics of the cubic nonlinear
  schr{\"o}dinger equation,'' {\em Physical Review Letters}, vol.~119, no.~3,
  p.~033901, 2017.

\bibitem{agrawal}
G.~Agrawal, {\em Nonlinear fiber optics}.
\newblock Academic Press, 5th~ed., 2013.

\bibitem{Lipton2015RNN}
Z.~C. Lipton, ``A critical review of recurrent neural networks for sequence
  learning,'' {\em Computing Research Repository}, vol.~abs/1506.00019, 2015.

\bibitem{hochreiter1997long}
S.~Hochreiter and J.~Schmidhuber, ``Long short-term memory,'' {\em Neural
  Computation}, vol.~9, no.~8, pp.~1735--1780, 1997.

\bibitem{Carleoreview2019}
G.~Carleo, I.~Cirac, K.~Cranmer, L.~Daudet, M.~Schuld, N.~Tishby,
  L.~Vogt-Maranto, and L.~Zdeborov\'a, ``Machine learning and the physical
  sciences,'' {\em Reviews of Modern Physics}, vol.~91, p.~045002, Dec 2019.

\bibitem{reviewRNN2019}
Y.~Yu, X.~Si, C.~Hu, and J.~Zhang, ``A review of recurrent neural networks:
  Lstm cells and network architectures,'' {\em Neural Computation}, vol.~31,
  pp.~1--36, 05 2019.

\bibitem{dudley2006supercontinuum}
J.~M. Dudley, G.~Genty, and S.~Coen, ``Supercontinuum generation in photonic
  crystal fiber,'' {\em Reviews of Modern Physics}, vol.~78, no.~4, p.~1135,
  2006.

\bibitem{farfan2018femtosecond}
C.~A. Farfan, J.~Epstein, and D.~B. Turner, ``Femtosecond pulse compression
  using a neural-network algorithm,'' {\em Optics Letters}, vol.~43, no.~20,
  pp.~5166--5169, 2018.

\bibitem{werbos1990backpropagation}
P.~J. Werbos, ``Backpropagation through time: What it does and how to do it,''
  {\em Proceedings of the IEEE}, vol.~78, no.~10, pp.~1550--1560, 1990.

\bibitem{chollet2015keras}
F.~Chollet {\em et~al.}, ``Keras.'' \url{https://keras.io}, 2015.

\bibitem{abadi2016tensorflow}
M.~Abadi, P.~Barham, J.~Chen, Z.~Chen, A.~Davis, J.~Dean, M.~Devin,
  S.~Ghemawat, G.~Irving, M.~Isard, {\em et~al.}, ``Tensorflow: A system for
  large-scale machine learning,'' in {\em 12th USENIX Symposium on Operating
  Systems Design and Implementation (OSDI 16)}, pp.~265--283, 2016.

\bibitem{tieleman2012lecture}
T.~Tieleman and G.~Hinton, ``Lecture 6.5-rmsprop: Divide the gradient by a
  running average of its recent magnitude,'' {\em COURSERA: Neural networks for
  machine learning}, vol.~4, no.~2, pp.~26--31, 2012.

\end{thebibliography}

\end{document}